# Optical analog of particle production in gravitational fields

Igor I. Smolyaninov

*Department of Electrical and Computer Engineering, University of Maryland, College Park, MD 20742, USA*

*e-mail: smoly@umd.edu*

**Strong enough electric field is predicted to spontaneously create electron-positron pairs in vacuum via Schwinger effect. A somewhat similar effect is predicted to occur in sufficiently strong inhomogeneous gravitational fields. However, due to necessity of very large field strength, these effects were never observed in the experiment. Here we demonstrate that optical analogs of very strong gravitational fields (up to ~$10^{24}$ g) and very strong gravitational field gradients (up to ~$10^{31}$ s$^{-2}$) may be created in electromagnetic metamaterial waveguides, leading to an optical analog of particle production in gravitational fields. Such waveguide geometries may also potentially be used to search for axion-like particles weakly interacting with electromagnetic field.**

One of the well-known predictions of quantum electrodynamics (QED) is that electron-positron pairs may be spontaneously created in vacuum in the presence of a very strong electric field. This prediction is called the Schwinger effect [1]. Schwinger pair production in a constant electric field takes place at a constant rate per unit volume $\Gamma$:

$$\Gamma = \frac{(eE)^2}{4\pi^3 c\hbar^2} \sum_{n=1}^{\infty} \frac{1}{n^2} e^{-\frac{\pi m^2 c^3 n}{eE\hbar}} \qquad (1)$$

where $m$ is the electron mass, and $E$ is the electric field strength. Since pair production is exponentially slow when the electric field strength is much below the Schwinger limit

$$E_s = \frac{m^2 c^3}{e\hbar} \approx 1.32 \times 10^{18} \text{ V/m,} \qquad (2)$$

Schwinger effect has never been observed in the experiment. For example, various proposals to observe Schwinger effect in a strong laser field so far fall short of the required electric field strength [2]. The electric field $E_s$ can be understood as the field which can pull a couple of virtual charged particles of mass $m$ out of quantum vacuum to a distance of the order of the Compton wavelength of the particle

$$\lambda_C = \frac{\hbar}{mc} \ , \qquad (3)$$

so that these particles become real. The corresponding condition for the electric field strength is

$$eE_s \lambda_C \sim mc^2 \qquad (4)$$

It was noted by several authors [3-5] that a somewhat similar effect of particle production must exist in strong enough inhomogeneous gravitational fields. However, in contrast to QED, the role of gravitational "charge" is played by the particle mass, which is always positive. As a result, a homogeneous constant gravitational field cannot separate particles in a virtual pair from each other. Nevertheless, the gravitational tidal forces (which cannot be transformed away through adequate coordinate transformations) may still be responsible for separation of the particles in a virtual pair [3,5]. For example, in the case of weak inhomogeneous gravitational field the local rate of massless particle production was calculated in [3] as

$$\Gamma = \frac{\alpha}{32\pi} C_{iklm} C^{iklm} \quad ,$$  (5)

where $C_{iklm}$ is the Weyl conformal tensor, and $\alpha$ is equal to 1/60 in the case of massless scalar particles, and 1/5 in the case of photons. Qualitatively, one must expect pair creation when the work exercised by tidal forces over the Compton wavelength of the particle $\lambda_C$ exceeds the rest mass of the particle pair [5]:

$$m\nabla G \lambda_C^2 \sim mc^2 \quad ,$$  (6)

where $\nabla G$ is the gravitational field gradient. Therefore, an order of magnitude of the "Schwinger gradient" of the gravitational field $\nabla G_S$, which is necessary for particle production may be estimated as

$$\nabla G_s = \frac{m^2 c^4}{\hbar^2}$$  (7)

If $m$ is assumed to be equal to the electron mass, $\nabla G_S \sim 6 \cdot 10^{41} s^{-2}$ is needed to observe the gravitational particle production. If existence of massive axion-like particles [6] weakly interacting with electromagnetic field is assumed (axions are considered to be the leading dark matter candidate [7]), the gravitational field gradients required to observe axion creation due to gravitational tidal forces becomes smaller. According to most estimates, the axion mass $m_a$ must fall in between 50 and 1,500 μeV [8], which means that gravitational field gradients of the order of $\nabla G_S \sim 5 \cdot 10^{21}$ - $10^{26} s^{-2}$ must be necessary for axion creation due to gravitational tidal forces. In any case, only in the early universe the gravitational fields were probably strong enough to lead to observable consequences of the gravitational particle production [3-5].

As noted in [5], the gravitational particle production is closely related to Hawking radiation and Unruh effect [9]. Unruh effect predicts that an accelerating object perceives its surroundings as a bath of thermal radiation, even if it accelerates in vacuum. Similar to the gravitational particle production, Unruh effect is also believed to be very difficult to observe in the experiment, since an observer accelerating at $g$=9.8 m/s$^2$ should see vacuum temperature of only $4 \cdot 10^{-20}$ K. Nevertheless, very recently it was noted that sufficiently strong accelerations for experimental observation of Unruh effect may be created in specially designed optical waveguides [10]. Using hyperbolic metamaterials [11] the upper limit on the effective accelerations may be pushed towards $\sim 10^{24} g$ [12]. Due to the Einstein equivalence principle, similar settings may also be used to study the physics of record high gravitational fields. By producing spatial variations of the effective acceleration, effective gravitational field gradients of the order of $\sim 10^{24} g / \lambda \sim 10^{31}$ s$^{-2}$ may be emulated in the metamaterial waveguides (where $\lambda$ is the photon wavelength). Based on the estimates above, such large magnitudes of the effective field gradient may lead to an optical analog of the gravitational particle production in electromagnetic metamaterials.

Let us demonstrate how an optical analog of a strong gravitational field may be created using electromagnetic metamaterials. The equations of electrodynamics in the presence of static gravitational field are identical to Maxwell equations in an effective medium in which $\varepsilon = \mu = g_{00}^{-1/2}$, where $g_{00}$ is the $tt$ component of the metric tensor [13]. It is convenient to describe effects arising in a homogeneous gravitational field (or in a reference frame moving with constant acceleration) using the Rindler metric [13]:

$$ds^2 = -\frac{G^2 x^2}{c^2} dt^2 + dx^2 + dy^2 + dz^2 \;, \tag{8}$$

where $G$ is the constant proper acceleration measured by a co-moving accelerometer. The analogy noted above indicates that very large constant gravitational field $G$ may be emulated using a metamaterial medium exhibiting the following coordinate dependencies of its dielectric permittivity $\varepsilon$ and magnetic permeability $\mu$:

$$\varepsilon = \mu = \frac{c^2}{Gx} \tag{9}$$

A typical order of magnitude of effective accelerations achievable in such an optical analog configuration may be estimated as at least

$$G_{eff} \sim \frac{c^2}{\lambda} \;, \tag{10}$$

which reaches $\sim 10^{22} g$ in the visible frequency range (assuming that optical experiments are conducted at $\lambda \sim 1 \mu$m). As noted above, even larger magnitudes of effective accelerations of the order of $\sim 10^{24} g$ appear to be achievable in some hyperbolic metamaterial geometries [12]. However, in order for this analogy to be precise, one must assume zero imaginary parts of $\varepsilon$ and $\mu$, which is not typically the case in electromagnetic metamaterials. On the other hand, this requirement may be met by incorporating an optically (or electronically) pumped gain medium into the metamaterial design, as described for example in [14]. External energy pumped into the metamaterial may enable particle creation due to either Unruh effect or strong tidal forces, which would be otherwise prohibited by energy conservation.

Photons in a waveguide behave as massive quasi-particles which may be characterized by both inertial and gravitational mass, which obey the Einstein equivalence principle [12,15,16]. Therefore, creation of such massive photons in a waveguide-based optical analog of strong gravitational fields may be considered as optical analog of gravitational particle production. The estimates made above strongly indicate that the order of magnitude of the analog gravity strength looks sufficient for experimental observation of this effect.

Let us consider an empty rectangular optical waveguide shown in Fig. 1(a), which walls are made of an ideal metal, and assume that this waveguide has constant dimensions ($d$ and $b$) in the $z$- and $y$- directions, respectively. The dispersion law of photons propagating inside this waveguide coincides with a dispersion law of massive quasi-particles:

$$\frac{\omega^2}{c^2} = k_x^2 + \frac{\pi^2 I^2}{d^2} + \frac{\pi^2 J^2}{b^2} \quad , \qquad (11)$$

where $k_x$ is the photon wave vector in the $x$-direction, $\omega$ is the photon frequency, and $I$ and $J$ are the mode numbers in the $z$- and $y$- directions, respectively. The effective inertial rest mass of the photon in the waveguide is

$$m_{eff} = \frac{\hbar \omega_{ij}}{c^2} = \frac{\hbar}{c} \left( \frac{\pi^2 I^2}{d^2} + \frac{\pi^2 J^2}{b^2} \right)^{1/2} \qquad (12)$$

It is equal to its effective gravitational mass [12,16]. Let us assume that this waveguide is either immersed in a constant gravitational field which is aligned along the x-direction, or (equivalently) is subjected to accelerated motion. As demonstrated above, since the gravitational field is static, this geometry may be represented by a static waveguide filled with a medium, in which $\varepsilon$ and $\mu$ gradually change as a function of x-coordinate (see Fig.1(a)). In the weak gravitational field limit

$$g_{00} \approx 1 + \frac{2\phi}{c^2} \quad , \tag{13}$$

where the gravitational potential $\phi = Gx$. Therefore, material parameters representing such a waveguide may be chosen so that $\varepsilon = \mu$, and the refractive index $n$ of the waveguide has a gradient in the $x$-direction:

$$n = (\varepsilon\mu)^{1/2} \approx 1 - \frac{Gx}{c^2} \tag{14}$$

This means that the "optical dimensions" of the waveguide in the transverse directions change as a function of the $x$-coordinate. Such a waveguide is called a tapered waveguide. Similar to massive bodies, photons in this waveguide moving against an external gravitational field eventually stop and turn around near the waveguide cut-off. The effective waveguide acceleration may be related to the refractive index gradient as

$$a = G = -c^2 \frac{dn}{dx} \tag{15}$$

In fact, as demonstrated in [12], the effect of spatial gradients of $\varepsilon$ and $\mu$ inside the waveguide and the effect of waveguide tapering (see Fig.1(b)) are similar, so that an effective waveguide acceleration may be calculated as

$$a = c_{gr} \frac{dc_{gr}}{dx} \quad , \tag{16}$$

where $c_{gr} = d\omega/dk_x$ is the group velocity of photons in the waveguide. If the gravitational field gradient $\nabla G$ needs to be emulated, in the first order approximation it may be obtained as

$$\nabla G = -c^2 \frac{d^2 n}{dx^2} \qquad (17)$$

Optical analog of the gravitational particle production in the waveguide geometries is illustrated in Fig.2. Real massive photons may be created from the virtual (evanescent) ones due to strong gradient of effective gravitational field in both the gradient index metamaterial waveguide geometry (Fig.2(a)) and the tapered waveguide geometry (Fig.2(b)). In fact, the latter configuration almost coincides with the well-known photon scanning tunneling microscopy (PSTM) geometry [17], which means that the optical analog of gravitational particle production should be relatively straightforward to observe and study, as illustrated in Fig.3.

Fig.3(a) shows a scanning electron microscope image of a PSTM optical fiber probe which was fabricated by overcoating a tapered optical fiber (its original shape is indicated by the continuous red lines in Fig.3(a)) with a thick aluminum layer, and cutting its end using focused ion beam milling [17]. A 100 nm open aperture (visible as a dark black circle) was left at the tip apex. The approximate positions of the waveguide cutoff at $\lambda_0$=632 nm and the Rindler horizon, as perceived by the photons passing through the cutoff region, are shown by the dashed lines. The effective acceleration and the approximate position $X$ of the Rindler horizon for these photons may be estimated using Eqs. (7) and (15) as

$$X \approx \frac{c^2}{a} \approx \frac{\lambda_0}{2n(db/dx)} \qquad (18)$$

leading to $a \sim 10^{22}\ g$ (note also that the photon acceleration in the waveguide is not constant and therefore no real horizon exists in this geometry). A 3.25x3.25 $\mu m^2$ PSTM image of a test sample prepared by milling 100 nm wide linear apertures through a 50 nm thick aluminum film deposited onto a glass slide surface is shown in Fig.3(b). This image was collected using the tapered waveguide shown in Fig.3(a), which was raster scanned over the sample surface. The sample was illuminated from the bottom with 632 nm light. This simple experiment demonstrates that real massive photons (contributing to the image in Fig.3(b)) may be created from the virtual (evanescent) ones. The only difference between this demonstration and the optical analog of gravitational particle production is that the photons contributing to the PSTM image are not created spontaneously. Indeed, in the experimental geometry shown in Fig.3(a) the effective gravitational field gradient estimated using Eqs.(17,18) is of the order of $\sim 10^{22} g/\lambda \sim 2 \cdot 10^{29}$ s$^{-2}$. It is of the same order of magnitude as the "Schwinger" gravitational field gradient $\nabla G_S$ determined by Eqs. (7) (using the effective photon mass defined by Eq.(12)).

Thus, the proper experimental geometries for the observation of optical analog of gravitational particle production may be summarized in Fig.4. In the gradient index metamaterial waveguide configuration depicted in Fig.4(a), optical pumping of the gain medium component of the metamaterial-filled waveguide (which is necessary to achieve Im($\varepsilon$)=Im($\mu$)=0 conditions) leads to directional flux of the generated "Schwinger" photons along the direction of the effective gravitational field $G$ and its gradient. These photons are created spontaneously through the volume of the metamaterial medium. The effective gravitational tidal forces must pull apart these virtual massive photons to a distance of the order of their Compton wavelength, at which point they become real. A similar tapered waveguide configuration is shown in Fig.4(b).

The optical analog of gravitational particle production may be observed directly via measurements of the directional flux of the "Schwinger photons". In the $\nabla G >> \nabla G_S$ limit the rate of "Schwinger photon" production in a waveguide may be estimated based on Eq.(1) as

$$\Gamma \approx V \frac{(m\lambda_C \nabla G)^2}{24\pi c\hbar^2} = V \frac{(\nabla G)^2}{24\pi c^3} \quad , \qquad (19)$$

where $V$ is the waveguide volume (in this estimate we have used the analogy between the $eE$ and $m\lambda_C \nabla G$ factors in the expressions for particle production rates due to strong electric field and due to strong gravitational field gradient, respectively). Note that Eq.(19) agrees with Eq.(5) above, which was derived in [3]. This effect appears to be quite pronounced since as estimated above, the order of magnitude of the effective gravitational field gradient may reach up to $\nabla G \sim 10^{31}$ s$^{-2}$. In principle, it could also be used to detect hypothesized axion-like particles which may be also created in addition to photons due to the optical analog of gravitational particle production. As we have discussed above, gravitational field gradients of the order of $\nabla G_S \sim 5 \cdot 10^{21}$ - $10^{26}$ s$^{-2}$ must be necessary for axion creation due to gravitational tidal forces, while the range of effective gravitational field gradients which may be created using the proposed electromagnetic waveguide geometries is considerably larger (up to $10^{31}$ s$^{-2}$). Moreover, while in typical "axion electrodynamics" models the axion field is relatively weakly coupled to the other electromagnetic degrees of freedom [6,7], this coupling is still believed to be strong enough, so that several microwave cavity-based "axion haloscopes" have been suggested to look for the hypothesized dark matter axions [18,19].

The equations of macroscopic axion electrodynamics [20] are typically introduced in such a way that the hypothetic axion contributions are incorporated into the macroscopic $D$ and $H$ fields:

$$\vec{D}_a = \vec{D} - g_{a\gamma\gamma}\left(\phi\vec{B}\right) , \tag{20}$$

$$\vec{H}_a = \vec{H} + g_{a\gamma\gamma}\left(\phi\vec{E}\right) , \tag{21}$$

where $\phi$ is the pseudo-scalar axion field and $g_{a\gamma\gamma}$ is the small axion-photon coupling constant. The resulting set of macroscopic Maxwell equations appears to be unchanged [20]:

$$\vec{\nabla} \cdot \vec{D}_a = \rho_f , \tag{22a}$$

$$\vec{\nabla} \cdot \vec{B} = 0 , \tag{22b}$$

$$\vec{\nabla} \times \vec{E} = -\frac{\partial \vec{B}}{\partial t} , \tag{22c}$$

$$\vec{\nabla} \times \vec{H}_a = \vec{J}_f + \frac{\partial \vec{D}_a}{\partial t} , \tag{22d}$$

where $\rho_f$ and $J_f$ are the densities of free charges and currents. In fact, several solid state systems, such as magnetoelectric antiferromagnet chromia [21], exhibit axion-like quasiparticles which indeed follow this description. If the metamaterial waveguide shown in Fig.4(a) is made using the magnetoelectric antiferromagnet chromia as one of its component, these axion-like quasiparticles will be also generated in addition to photons inside the waveguide due to the optical analog of gravitational particle production. Since the equations of macroscopic electrodynamics are known to be identical to the equations of electrodynamics in vacuum in the presence of gravitational field (see [13] and a very detailed discussion in [22]), emission of "real" vacuum axions due to the optical analog of gravitational particle production must be also expected if strong enough effective gravitational field gradient is created.

Indeed, as demonstrated by Capolupo et al. [23], in the quantum field theoretical framework axions and photons exhibit considerable mixing, which is somewhat similar to the mixing and oscillations of the neutrino flavors. Following [23], let us consider either one of the waveguides shown in Fig.4(a,b), which is subjected to a DC magnetic field $B_0$ directed along the $y$ coordinate. For the guided photon propagation along the $x$ direction, the photon polarizations decouple, and the axion-photon mixing matrix for the state of polarization parallel to $B_0$ may be obtained as

$$M = -\frac{1}{2\omega}\begin{pmatrix} m_{eff}^2 & -g_{a\gamma\gamma}\omega B_0 \\ -g_{a\gamma\gamma}\omega B_0 & m_a^2 \end{pmatrix}, \qquad (23)$$

where $m_{eff}$ is the effective photon mass in the waveguide, and the natural units are used for $B_0$ and all the other parameters. The mixing matrix can be diagonalized by mixing of the photon $\gamma$ and axion $\phi$ fields:

$$\begin{pmatrix} \gamma' \\ \phi' \end{pmatrix} = \begin{pmatrix} \cos\theta & \sin\theta \\ -\sin\theta & \cos\theta \end{pmatrix}\begin{pmatrix} \gamma \\ \phi \end{pmatrix}, \qquad (24)$$

where

$$\theta = \frac{1}{2}\arctan\left(\frac{2g_{a\gamma\gamma}\omega B_0}{m_a^2 - m_{eff}^2}\right) \qquad (25)$$

is the mixing angle, $\gamma$ and $\phi$ are the fields associated with "mixed" particles, and $\gamma'$ and $\phi'$ are the "free" fields with definite masses. While in free space $m_{eff}=0$ and the mixing angle $\theta$ is very small, in the waveguide geometry the effective mass of the photon is defined by Eq.(12). If $m_{eff}=m_a$ conditions are met in the waveguide, the axion-photon mixing is greatly enhanced, so that similar to photons, axions will experience the optical analog of the gravitational field $G$ defined by Eq.(9). As was noted in [23], the $m_{eff}=m_a$ conditions in free space may be also met in the presence of plasma, so that $m_{eff}$ would be equal to the plasma frequency. However, the lifetime of photons in plasma is rather short, so that axion-photon mixing defined by Eq.(25) will be limited by the imaginary part of $m_{eff}$. On the other hand, the lifetime of photons in optical fibers is extremely long. The typical photon propagation length in optical fibers reaches $L \sim 100$ km. Therefore, as illustrated in Fig.5(b), the axion-photon mixing in optical fibers may become very large. The absolute value of the axion-photon mixing angle was calculated using Eq.(25) assuming $\mathrm{Im}(m_{eff}) \sim 1/L$ in the geometry shown schematically in Fig.5(a). Based on Eq.(12), it was assumed that an optical fiber with a core diameter $d \sim 50$ μm is used to achieve $m_{eff}=m_a$ conditions at light frequency $\omega \sim 2\mathrm{eV}$ and the fiber is subjected to perpendicular magnetic field $B_0$. The fiber is gradually tapered, which according to Eq.(12) leads to gradual changes of $m_{eff}$. It is clear from Fig.5(b) that gradual tapering of such a waveguide must lead to a very strong increase of the axion-photon coupling in the fiber taper region where $m_{eff}=m_a$. Experimental observations of the enhanced axion-photon mixing in this geometry may be performed using the "light shining through the wall" [24] approach. Indeed, the photon contribution to the gravitational Schwinger effect particle flux defined by Eq.(19) may be eliminated by closing the waveguide on both sides with sufficiently opaque metal layers, thus making it into a tapered waveguide cavity. However, weakly interacting vacuum axions (if any) created due to optical analog of gravitational particle production may still be able to leave the cavity.

Similar to the "light shining through the wall" experiments, this axion flux may be detected using an adjacent similar tapered waveguide. As mentioned above, application of a very strong DC magnetic field would considerably enhance axion-photon mixing.

In conclusion, the record high accelerations up to $10^{24}$ g and their extreme gradients up to $10^{31}$ $s^{-2}$, which may be created using metamaterial waveguides in terrestrial laboratories, appear to enable experimental studies of the gravitational particle production. Such experiments may also be potentially used to search for axion-like particles weakly interacting with conventional matter. While there could be no spontaneous particle production in passive optical media, the optical analog of gravitational particle production discussed in this paper is made possible by incorporation of gain media into various waveguide geometries (as illustrated in Fig. 4). In addition, as pointed out in [25], the presence of horizons is not relevant for particle production by tidal forces, so the absence of analog horizons in the considered waveguide geometries does not preclude the optical analogs of gravitational particle production. We should also point out that similar to detailed consideration in ref. [26], the effects of dispersion in the optical waveguide geometries also enable particle production in the absence of analog horizons.

**Figure Captions**

**Figure 1.** Photon in a waveguide behaves as a massive quasi-particle: (a) Similar to massive bodies, photons in a waveguide moving against an external gravitational field eventually stop and turn around near the waveguide cut-off. Gradient of the effective refractive index $n$ in the waveguide is illustrated by shading. (b) The effect of external gravitational field on a photon in a waveguide may be emulated by waveguide tapering, which also leads to accelerated motion of massive photons.

**Figure 2.** Creation of real massive photons from the virtual (evanescent) ones due to strong gradient of effective gravitational field (a) in a gradient index metamaterial waveguide, and (b) in a tapered waveguide. The latter configuration almost coincides with the photon scanning tunneling microscopy (PSTM) geometry.

**Figure 3**. (a) SEM photograph of a PSTM optical fiber probe which was fabricated by overcoating a tapered optical fiber (indicated by the continuous red lines) with a thick aluminum layer, and cutting its end using focused ion beam milling [17]. A 100 nm open aperture (visible as a dark black circle) was left at the tip apex. The approximate positions of the waveguide cutoff at $\lambda_0$=632 nm and the Rindler horizon, as perceived by the photons passing through the cutoff region, are shown by the dashed lines. Note that the photon acceleration is not constant. (b) 3.25x3.25 $\mu m^2$ PSTM image of a test sample prepared by milling 100 nm wide linear apertures through a 50 nm thick aluminum film deposited onto a glass slide surface. The image was collected using a probe shown in (a), which was raster scanned over the sample surface. The sample was illuminated from the bottom with 632 nm light.

**Figure 4**. (a) Schematic geometry of the optical analog of gravitational particle production in the gradient index metamaterial waveguide configuration. Optical

pumping of the gain medium component of the metamaterial-filled waveguide, which is necessary to achieve Im($\varepsilon$)=Im($\mu$)=0 conditions, leads to directional flux of the generated "Schwinger" photons. (b) Schematic geometry of a similar tapered waveguide configuration.

**Figure 5**. (a) Schematic geometry of a tapered optical fiber which may achieve $m_{eff}=m_a$ conditions at light frequency $\omega$~2eV. The tapered fiber is subjected to perpendicular DC magnetic field $B_0$. (b) The absolute value of the axion-photon mixing angle $\theta$ calculated using Eq.(25) assuming Im($m_{eff}$)~$1/L=1/100\ km$ in the geometry shown schematically in (a).

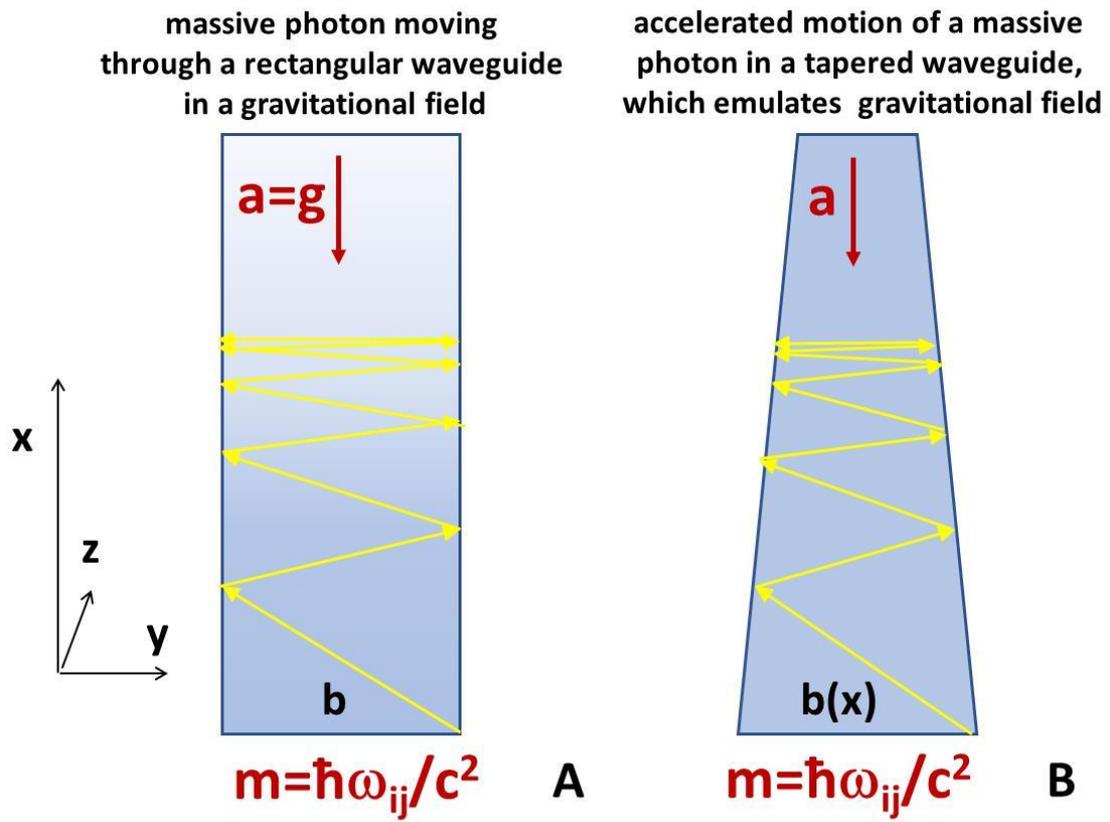

Fig. 1

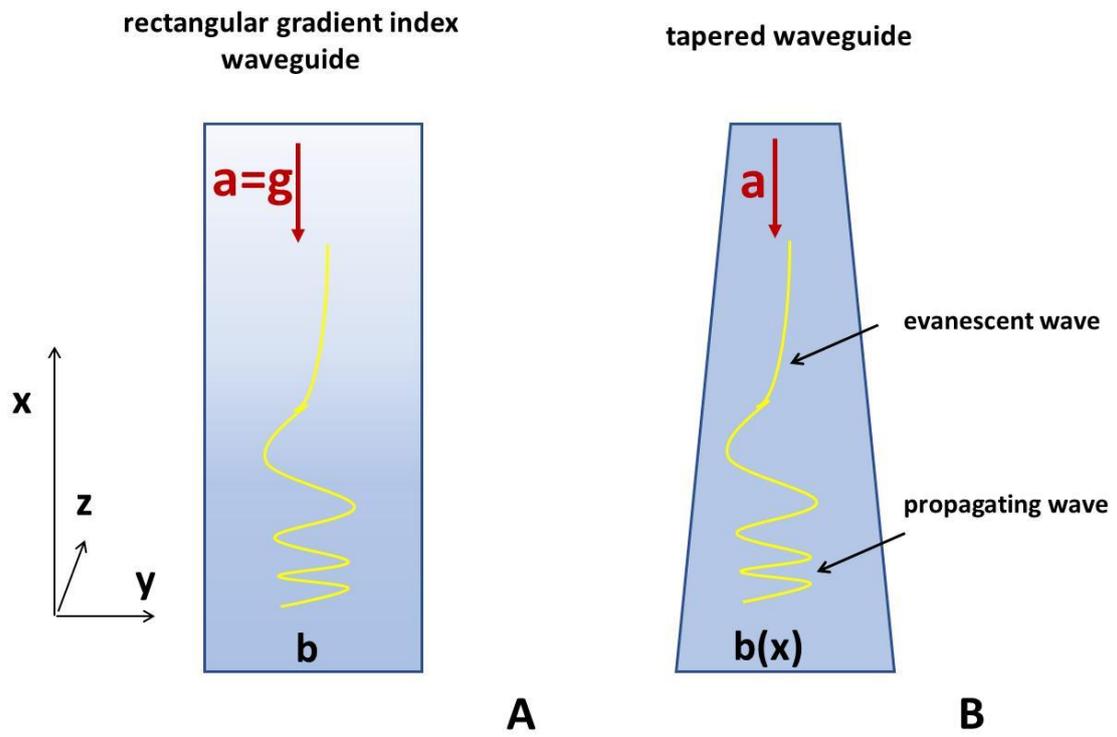

Fig. 2

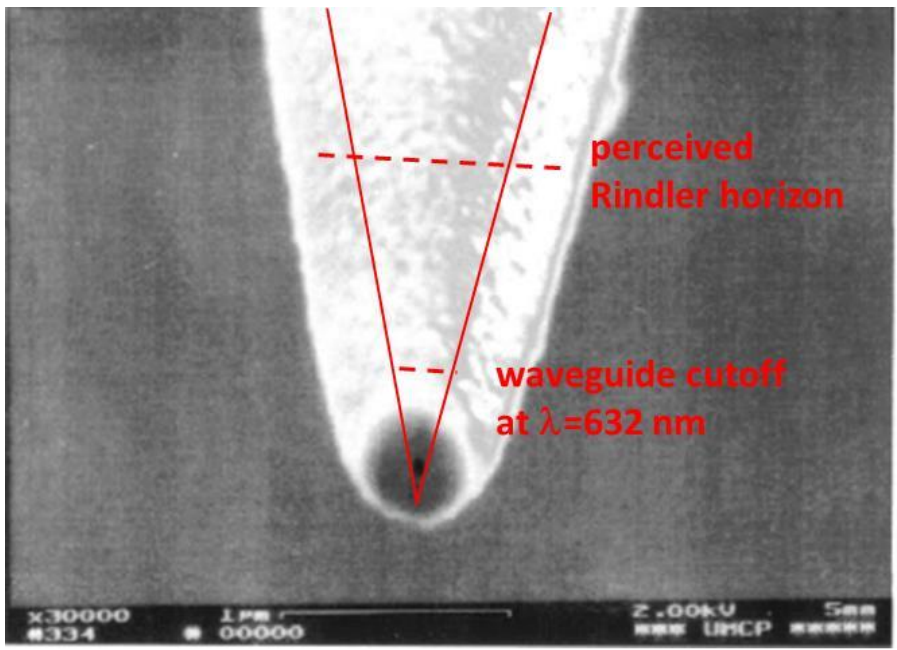

(a)

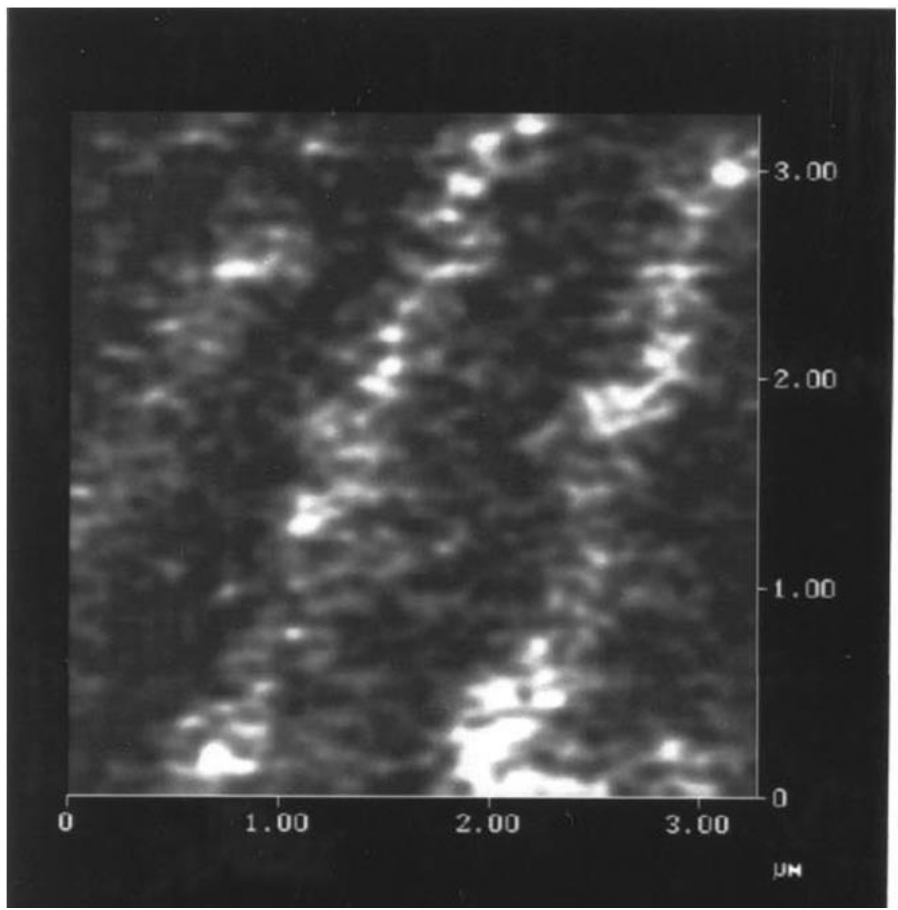

(b)

Fig. 3

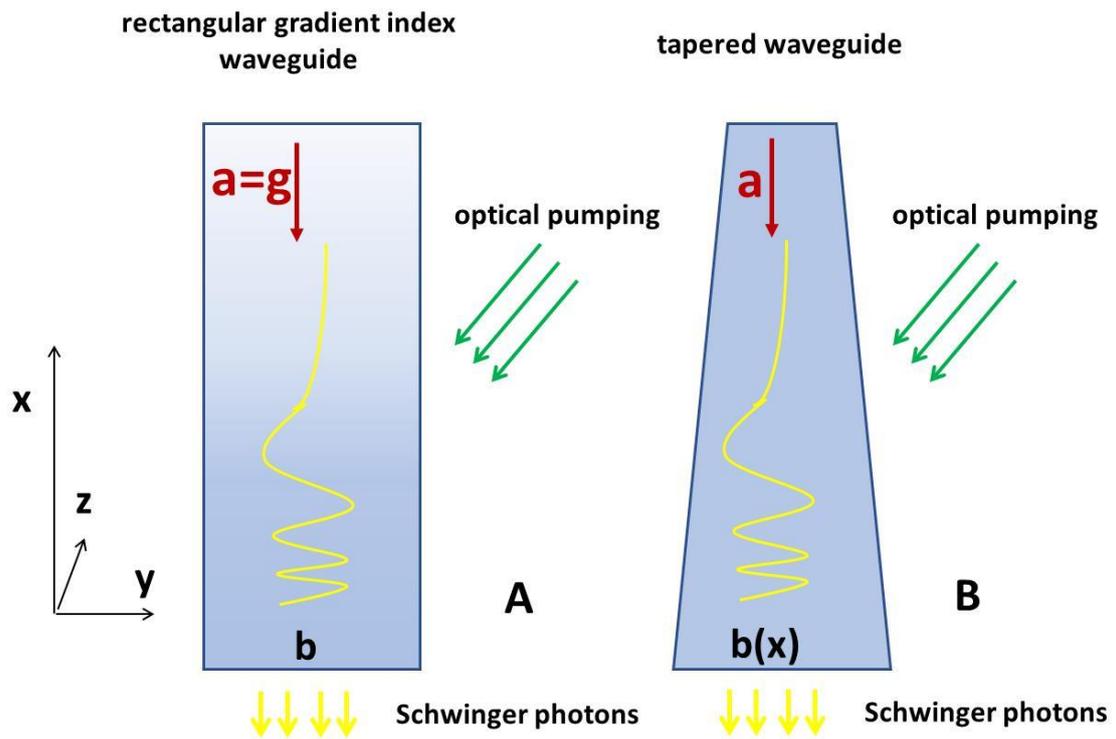

Fig. 4

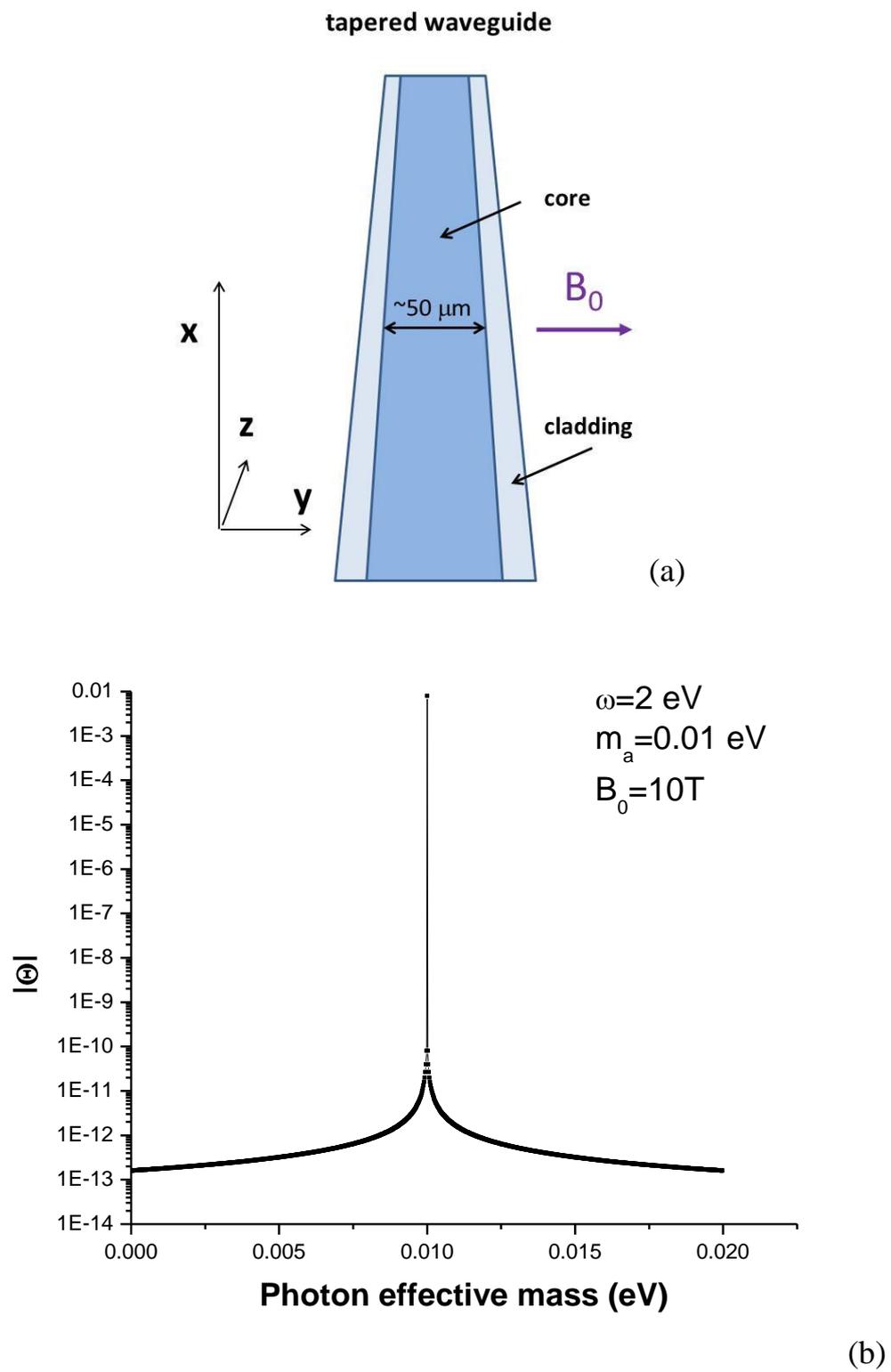

Fig. 5